\pdfoutput=1 
\newcommand{\arxiv}[2]{{\href{http://www.arxiv.org/abs/#1}{\tt arXiv:#1}} [#2]}
\documentclass{JINST}
\usepackage{subfigure}

\title{Massively Parallel Computing at the Large Hadron Collider up to the HL-LHC}

\author{P. Lujan$^a$\thanks{Corresponding author.}~
and V. Halyo$^a$\\
\llap{$^a$}Princeton University, Department of Physics,\\
  Princeton, NJ 08544, USA\\
E-mail: \email{plujan@princeton.edu}}

\abstract{As the Large Hadron Collider (LHC) continues its upward progression in energy and luminosity towards the
planned High-Luminosity LHC (HL-LHC) in 2025, the challenges of the experiments in processing increasingly
complex events will also continue to increase. Improvements in computing technologies and algorithms will be a
key part of the advances necessary to meet this challenge. Parallel computing techniques, especially those
using massively parallel computing (MPC), promise to be a significant part of this effort. In these
proceedings, we discuss these algorithms in the specific context of a particularly important problem: the
reconstruction of charged particle tracks in the trigger algorithms in an experiment, in which high computing
performance is critical for executing the track reconstruction in the available time. We discuss some areas
where parallel computing has already shown benefits to the LHC experiments, and also demonstrate how a
MPC--based trigger at the CMS experiment could not only improve performance, but also extend the reach of the
CMS trigger system to capture events which are currently not practical to reconstruct at the trigger level.}


\begin{document}

\section{Introduction}\label{sec:introduction}

The Large Hadron Collider (LHC) has already produced dramatic physics results in its ``Run 1'' period
(2010--2012), principally with the discovery of the Standard Model (SM) Higgs boson at a mass of approximately
125~GeV/$c^2$~\cite{Aad:2012tfa,Chatrchyan:2012ufa}, operating at a center-of-mass energy of $\sqrt{s} =
7$--$8$ TeV and typical luminosities of approximately $5 \times 10^{33}$ cm$^{-2}$ s$^{-1}$. Once the LHC
reaches its design energy of $\sqrt{s} = 14$ TeV, currently scheduled to happen during Run 2 (2015-2018),
however, the main gains in the performance of the accelerator will be realized by increasing the luminosity
over the course of the LHC lifetime, currently planned to culminate in the High-Luminosity LHC (HL-LHC) run
starting in 2025, with an expected luminosity of $5 \times 10^{34}$ cm$^{-2}$ s$^{-1}$. The HL-LHC period is
expected to see major upgrades to all LHC detectors to increase the physics capabilities, replace
radiation-damaged components, and deal with the increased luminosity.

As the luminosity continues to increase, the role of computing in processing the increasing volumes of data,
both online and offline, will take on greater importance. One promising technique which has drawn increased
attention in recent years is the use of parallel computing, especially since many computing problems in
high-energy physics are naturally well-suited to parallel processing. Products such as the NVIDIA Tesla,
which use general-purpose GPU (GPGPU) computing, and the Intel Xeon Phi, which is based on a Massively
Integrated Core (MIC) architecture, offer relatively cost-efficient and easy-to-integrate ways to add parallel
computing capabilities to existing computing facilities.

In this paper, we examine present and future uses of massively parallel computing techniques at the LHC
experiments, with a particular focus on the specific problem of track reconstruction in the online trigger
system. This is an area where high-performance computing is especially critical, as the trigger must make a
decision on an event in a very short time interval, and affects the potential physics reach of the detector.
Normally, compromises are necessary in the track reconstruction algorithms in order to improve their
performance to the required levels, which means that the trigger may be unable to recognize and capture
certain types of physics events. Advances in the computing techniques available at this level could thus allow
searches for entirely new physics which are not possible with current methods. We present a specific new
physics model where long-lived particles are produced which decay at a significant distance from the original
interaction point (but still within the tracking volume of the detector) and discuss the implications of a
parallel computing-based track reconstruction for the trigger in this model.

\section{From the LHC to the HL-LHC}\label{sec:lhc}

The LHC was designed to collide two proton beams at a center-of-mass energy of $\sqrt{s} = 14$ TeV, although it
has not yet operated at that energy. In the initial run of the LHC (``Run 1''), the collider operated at an
energy of $\sqrt{s} = 7$ TeV in 2010--2011, increasing to $\sqrt{s} = 8$ TeV in 2012. For the LHC restart in
2015, the LHC will operate initially at $\sqrt{s} = 13$ TeV, expected to rise over the course of ``Run 2''
(lasting from 2015-2018) to the design energy of $\sqrt{s} = 14$ TeV.

However, once the design energy has been attained, any further gains in accelerator performance must come from
increasing the luminosity. Partially this will be achieved by decreasing the bunch spacing of 50~ns in Run 1
to the design value of 25~ns, but beyond this, increased luminosity must also mean an increased number of
interactions per bunch crossing (pileup). The number of charged particle tracks and hence hits in the detector
will increase with the pileup, thus making events increasingly difficult to reconstruct in higher-luminosity
environments. The luminosity is planned to increase over the course of the LHC run, culminating in the
High-Luminosity LHC (HL-LHC) starting in 2025, with an expected luminosity of $5 \times 10^{34}$ cm$^{-2}$
s$^{-1}$, a full order of magnitude greater than the luminosity seen in Run 1. Table~\ref{tab:lhclumi}
summarizes the increase in energy, luminosity, and pileup expected over the course of the LHC run to the
HL-LHC period.

\begin{table}[tbp]
\caption{Energy, luminosity, and pileup at the LHC from Run 1 through the HL-LHC period. All numbers (except
for Run 1) are expected values, and hence approximate.}
\label{tab:lhclumi}
\smallskip
\centering
\begin{tabular}{|r|c|c|c|c|c|c|}
\hline
Period & Dates & $\sqrt{s}$ & Inst. lumi        & Bunch & Pileup & Total deliv. \\
       &       & (TeV)      & (cm$^{-2}$ s$^{-1}$)& spacing (ns) & & lumi (fb$^{-1}$) \\ 
\hline
Run 1 & 2010--2012 & 7--8 & $5 \times 10^{33}$ & 50 & 25 & 30 \\
Run 2 & 2015--2018 & 13--14 & $1.5 \times 10^{34}$ & 25 & 40 & 100 \\
Run 3 & 2020--2022 & 14 & $2 \times 10^{34}$ & 25 & 60 & 300 \\
HL-LHC & 2025-- & 14 & $5 \times 10^{34}$ & 25 & 130 & 3000 \\
\hline
\end{tabular}
\end{table}

Dealing with this increased luminosity will require substantial upgrades to the computing systems, both online
and offline, in order to be able to handle both the increased volume of data and the increased complexity of
this data. It is important that we develop new techniques in order to best handle this increased data, so we
can best take advantage of the power of the HL-LHC.

\section{Parallel Computing in Particle Physics}\label{sec:parallel}

In recent years, the increase in clock speed, which has been a strong driver of improved CPU performance, has
come nearly to a halt, as various intrinsic limitations of semiconductor devices have made further progress
along these lines difficult without a fundamental breakthrough. As a result, attention has shifted towards
possibilities in parallel processing, including multi-processor systems, multi-core processors, and
general-purpose computing using Graphical Processing Units (GPUs). These advances are naturally of interest to
the particle physics community, not simply because of the promise of increased performance, but also because
many computing problems in particle physics are naturally parallelizable and are so well-suited to approaches
using these new technologies.

General-purpose GPU (GPGPU) computing involves using the stream processors in a GPU to perform general numeric
computations. In its infancy, this required highly specialized code using graphics APIs such as DirectX or
OpenGL; however, recent developments have made GPGPU programming much more accessible. NVIDIA's introduction
of CUDA in 2007~\cite{cuda} provided a framework for developers to create GPU code using only a few additions
to standard C++ code, and in recent years a variety of standards (OpenMP, OpenACC, OpenCL, etc.) have emerged
to create an environment in which users can develop parallelized programs with a minimum of additional
effort. Of course, since the underlying architecture of a GPU is quite different from a standard CPU,
developers must think carefully about issues such as the way in which memory is accessed in order to obtain
the optimal performance from such GPGPU tools. Nevertheless, the barriers to using GPUs for all sorts of
general-purpose computing problems have never been lower. Similarly, products such as the NVIDIA Tesla line
provide a convenient solution for easily adding GPGPU computing capabilities to existing systems.

Another approach to highly parallel computing is the Massively Integrated Core (MIC) model, in which a very
large number of processor cores are available in a single chip. An example of this technology is the Intel
Xeon Phi, which uses a number of x86-based processor cores and is also available as a simple coprocessor
solution which can be easily added to supplement the power of an existing system.

Other approaches, such as using custom FPGAs to perform parallel tasks in, for example, reconstruction of
tracks in a collision event, are also being studied in the various experiments at the LHC. While these
approaches tend to require more specialized hardware and software than the GPGPU and MIC alternatives,
they offer the possibility of very high performance. The examples we discuss below do not use FPGA computing
specifically, but it is also an approach that shows much promise for the LHC experiments.

\section{Track Reconstruction and Triggering}\label{sec:tracking}

One particularly computing-intensive task in event reconstruction is the \textbf{tracking}: that is, the
reconstruction of charged particle tracks given the hits in the detector. Tracking is usually performed using
a combinatorial track finder (CTF) algorithm. As an example, consider the CMS tracking
algorithm~\cite{Chatrchyan:2014fea}. At CMS, the tracking is run iteratively: in earlier passes, tracks that
are easier to reconstruct (higher $p_T$ tracks originating from close to the interaction point) are found;
then, the hits corresponding to these tracks are removed, and further passes are run to look for more
difficult to reconstruct tracks. Each pass consists of four steps: a seeding step, the track finding, track
fitting, and selection.

In the seeding step, seeds are constructed either from triplets of hits or two hits and a primary vertex
constraint. In the track finding step, a trajectory is constructed and propagated through the detector; at
each layer, hits compatible with the trajectory on that layer are searched for and, if found, attached to the
candidate trajectory. In the case where more than one hit is compatible with the propagated trajectory, all
candidates are considered. Once the propagation through the detector is complete, the collection of candidates
is cleaned to remove duplicate tracks or tracks that share a large number of hits, selecting the best
candidate. Next, a Kalman fit is performed over the full track trajectory to obtain the best estimate of the
track parameters over the full course of the trajectory, first with a forward step running from the innermost
hits to the outermost, and then with a smoothing stage running backwards to use the information obtained from
the later hits to obtain a better hit for the inner hits. Finally, track selection requirements are applied to
select good-quality tracks.

As the number of pp interactions in an event increases, and thus the number of charged particle tracks and
hits in the tracker, the work that must be performed by a CTF-based algorithm increases exponentially, due to
the increasing number of possible combinations. As implemented in CMS (and similarly in other experiments),
the track reconstruction applies a variety of techniques to reduce the computing demand (such as the iterative
tracking mentioned above), but significant advances in performance will be required to keep up with the
tracking requirements as the LHC moves towards the HL-LHC era.

The problem of tracking reconstruction is particularly acute in the context of the online trigger. The LHC
delivers collisions to the experiments at a rate of 40 MHz, but the limitations of the data storage and
processing systems mean that this rate must be reduced to $\sim100$ Hz before final storage. This reduction
is accomplished by the online trigger, which typically consists of two parts: a hardware-based level 1
(L1) trigger which looks for interesting features in parts of the detector that can be read out quickly,
reducing the rate to approximately 100 kHz, and a software-based high-level trigger (HLT) which performs a
more complete reconstruction of the event to look for events containing physics signals of interest. Because
of the severe time limitations imposed on the HLT reconstruction, although the algorithms used are similar to
those offline, they impose more stringent limitations on what can be reconstructed. For example, at CMS, the
CTF algorithm is used for both online and offline track reconstruction; however, in the HLT, the track
reconstruction is only performed locally in regions near objects of interest identified by the L1 trigger, and
a pass to reconstruct tracks significantly displaced from the interaction point is not performed, as the
computational demands of such reconstruction exceed what is currently available to the HLT.

While every effort is of course made to ensure that the algorithms used at the HLT retain as much efficiency
as possible for the signatures of events that we expect to observe at an experiment like CMS, it is inevitable
that the HLT will not be able to consider every possibility. In particular, unusual exotic models can produce
events for which the current HLT is very inefficient. Thus, improved computing techniques do not merely give
us a way to process the more complex events expected in higher-luminosity running at the LHC, but also open
the possibility of searches for new physics which are simply not possible at all with the current trigger
algorithms. In Section~\ref{sec:hough}, we show an example of how a MPC-based algorithm can detect a new
physics model currently inaccessible to CMS.

\section{GPU Acceleration of the ALICE HLT Track Reconstruction}

As an example of parallel computing technologies that have already been implemented at the LHC experiments, we
consider the HLT used in the ALICE experiment, where GPUs have been implemented for reconstruction of
tracks in the central Time Projection Chamber (TPC) tracker~\cite{Rohr:2012nf}. The ALICE HLT track
reconstruction proceeds in two steps: first, a combinatorial cellular automaton technique is used to create
``tracklets'' by, for each hit, searching for hits in the adjacent rows which are compatible with a
straight-line fit. Then, a Kalman filter technique is used to obtain the trajectory for a tracklet and
propagate it through the detector to attach additional compatible hits. This naturally lends itself to a high
degree of parallelization, as each cluster or candidate track can be processed independently from the others.

\begin{figure}[tbp]
\centering
\includegraphics[width=.8\textwidth]{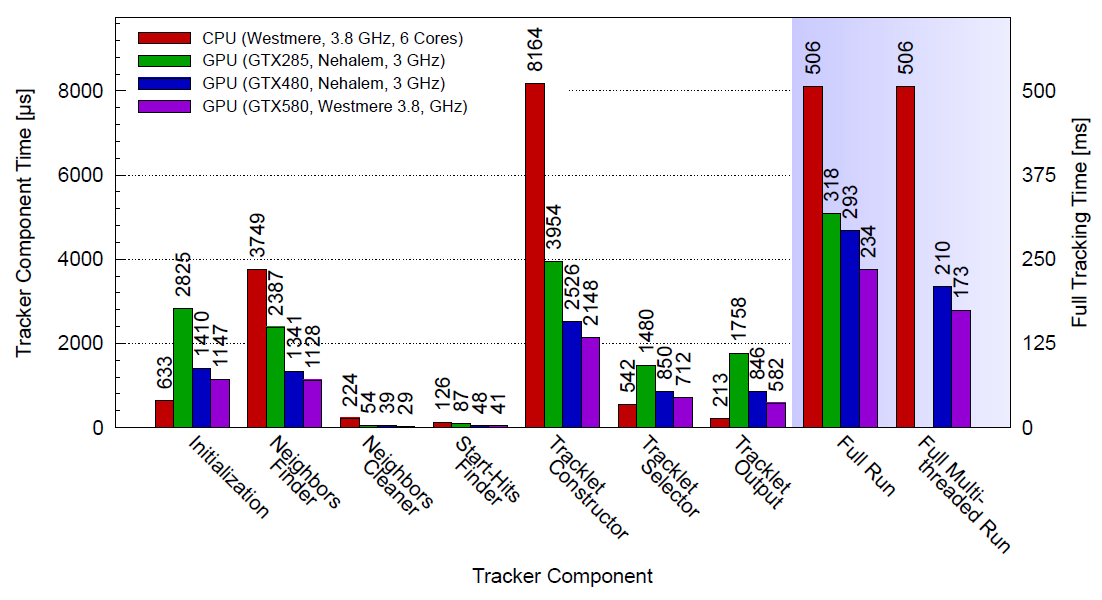}
\caption{Performance of the ALICE HLT TPC track reconstruction using the CPU-based algorithm (red bars) and
the GPU-based tracking algorithm for three different GPUs (green, red and purple). From~\cite{Rohr:2012nf}.}
\label{fig:AliceGPU}
\end{figure}

The GPU-based tracker was initially implemented using NVIDIA GTX 295 GPUs, and later upgraded to NVIDIA GTX
480 GPUs. In both cases these are standard commercially-available graphics cards. In the first implementation,
it was discovered that the GPU was often sitting idle while waiting for the CPUs to pre- and post-process the
events (these stages, being memory-bound, were unable to benefit from GPU acceleration). By developing a
multithreaded version, the GPU time could be used more efficiently and speed up the overall process. This
illustrates how careful attention to the various I/O demands of a process is often necessary to obtain the
best performance of a parallel implementation. Figure~\ref{fig:AliceGPU} shows the performance of the
GPU-based tracker on various different GPU/CPU combinations compared to the CPU-based
implementation~\cite{Rohr:2012nf}.

Overall, the GPU-based tracking algorithm showed an increase in performance by a factor of three, and since
the GPUs used in the implementation were substantially cheaper than the CPUs, the cost per performance was
even more favorable for the GPU implementation. It operated successfully through the heavy-ion runs in 2010,
2011, and 2012.

\section{Using Parallel Computing to Search for New Physics}\label{sec:hough}

The above example shows how parallel computing can substantially increase the performance of the track
reconstruction at a relatively low cost. However, the even more attractive benefit of using MPC technology is
that it can enable use of new algorithms which allow us to search for physics that is inaccessible to us with
the current track reconstruction in the HLT.

\subsection{Physics Motivation}

As an example, consider an event signature where tracks (either leptons or jets) are produced at a secondary
vertex at a significant displacement from the primary vertex (much longer than $b$-hadron decays, but still
within the tracker volume), due to the production and decay of some exotic long-lived neutral particle. Such a
signature would be a clear and unambiguous model of new physics, and is motivated by a wide variety of various
theoretical models~\cite{Halyo:2013yfa}, such as hidden valley models with long-lived neutral particles,
R-parity-violating supersymmetric (SUSY) models with long-lived neutralinos, split SUSY models with long-lived
gluinos, displaced black holes, boosted jets, or Z' models with long-lived neutrinos.

Searches have been performed at ATLAS~\cite{ATLAS:2012av,Aad:2012zx,Aad:2014yea} and
CMS~\cite{Chatrchyan:2012cg,CMS:2014hka} for such models, but the constraints imposed by the trigger system
limit their reach. For example, CMS has published a search for long-lived neutral particles which decay into
displaced leptons~\cite{CMS:2014hka}, covering two different possible signal models: one with a non-Standard
Model Higgs boson which decays into two long-lived neutral bosons, each of which decays into a pair of
leptons, and one where a squark decays into a long-lived neutralino, which then undergoes an
R-parity-violating decay into two leptons and a neutrino. However, because of the lepton momentum requirements
in the trigger, the resulting efficiency is very low for masses close to the SM Higgs mass of 125 GeV/$c^2$,
so that limits are best only for higher Higgs masses. An ideal solution would be a trigger that can look
specifically for displaced tracks meeting at a vertex, but using the existing CTF algorithm, such an algorithm
would be prohibitively computationally expensive, since the number of possible combinations allowed if one
allows significant displacement from the interaction point grows very rapidly.

\begin{figure}[tbp]
\centering
\includegraphics[width=.4\textwidth]{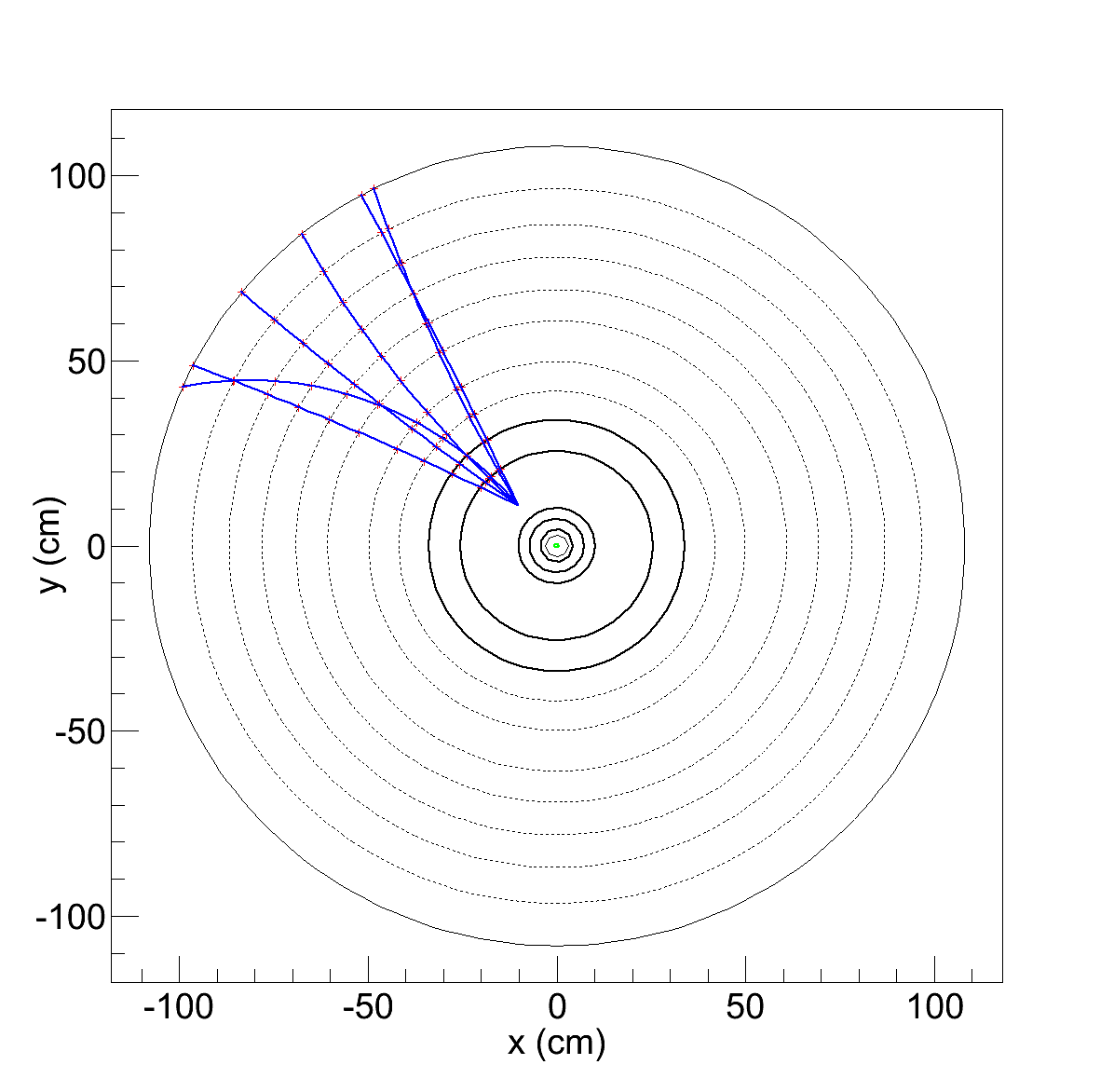}
\caption{A schematic of an event with a displaced jet in the CMS tracker. The layers shown in bold are the
layers used for seeding in the track reconstruction algorithms in the CMS HLT, and the green circle
illustrates the maximum transverse impact parameter (0.5 cm) within which a track can be reconstructed.}
\label{fig:DisplacedJet}
\end{figure}

Figure~\ref{fig:DisplacedJet} illustrates an example of such a displaced event in a schematic of the CMS
tracker. The bold layers are the tracker layers that are used for seeding, and the green circle in the center
shows the maximum transverse impact parameter $d_0$ of a track that can be reconstructed in the CMS HLT. It is clear
that in events such as these, the tracks can be missed entirely by the track reconstruction at the HLT level;
thus, we have to rely on less-efficient alternative triggering strategies.

\subsection{New Tracking Algorithms}

However, instead of using a traditional CTF-based approach, we can approach the problem of track
reconstruction by using a Hough transform algorithm. The Hough transform was originally developed for
machine-vision applications~\cite{bib:hough}, but has attracted a great deal of interest from the particle
physics community in recent years for its applicability to the problem of track reconstruction. In contrast to
an algorithm like CTF, which finds tracks one at a time, the Hough transform is a holistic approach which can
reconstruct all the tracks in an event in a single pass. The Hough transform operates by developing a
parameterization of the features being searched for (in this case, tracks) and constructing a parameter
space. Each hit in the detector then corresponds to a set of points in the parameter space representing tracks
passing through that point. After all hits are thus transformed, maxima in the parameter space correspond to
tracks which pass through many hits in the detector -- that is, the desired reconstructed tracks.

The Hough transform is highly parallelizable, as the transformation of each hit into the parameter space can
be done completely independently; only the final reduce step of combining and looking for maxima must be done
serially, and these steps are relatively simple computationally. The transformation into parameter space can
also naturally take into account the varying resolution of the hits observed in the detector.

\begin{figure}[tbp]
\begin{center}
\subfigure[]{\includegraphics[width=0.32\textwidth]{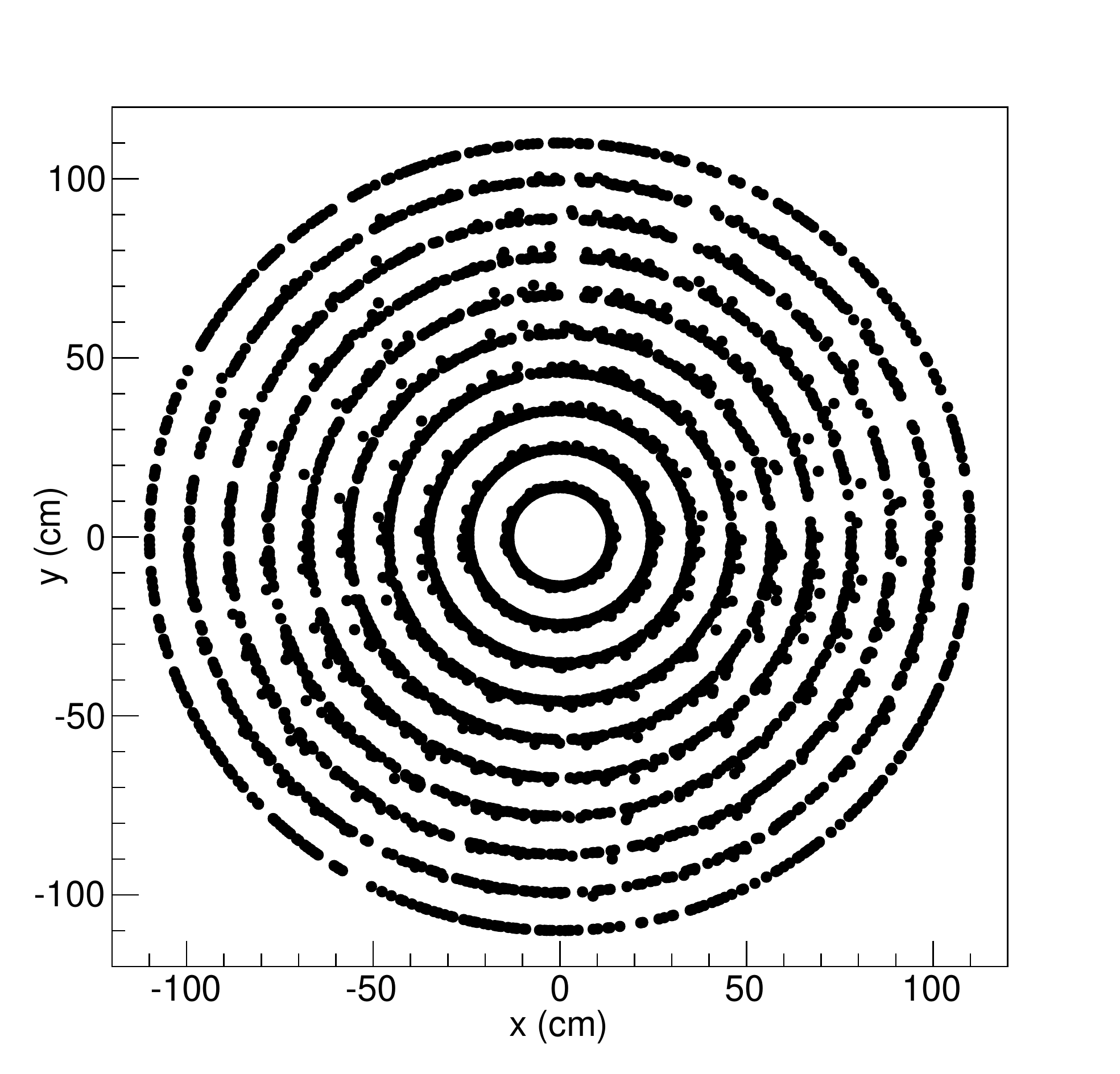} \label{fig:hits}}
\subfigure[]{\includegraphics[width=0.30\textwidth]{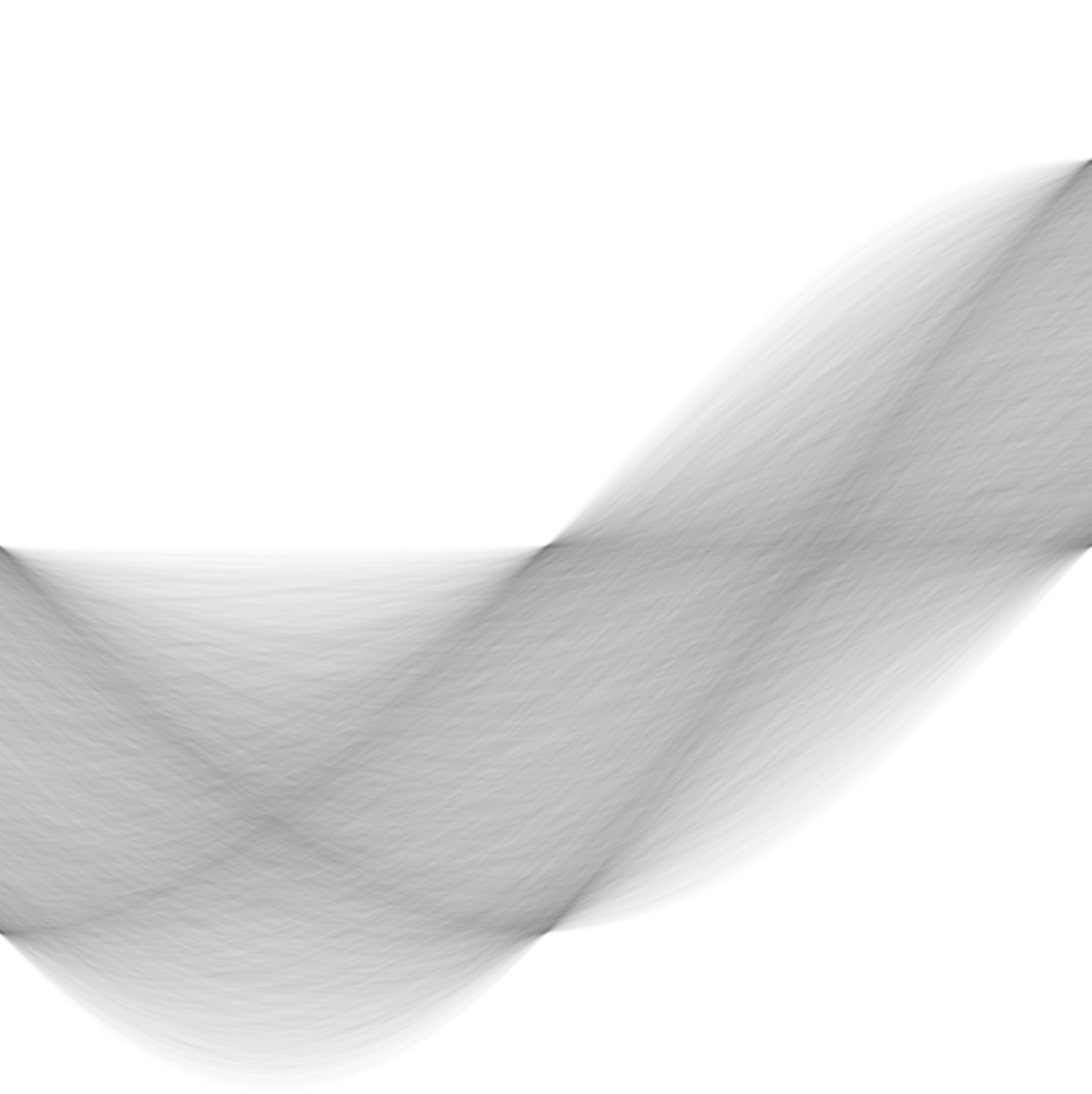} \label{fig:accumulator}}
\subfigure[]{\includegraphics[width=0.32\textwidth]{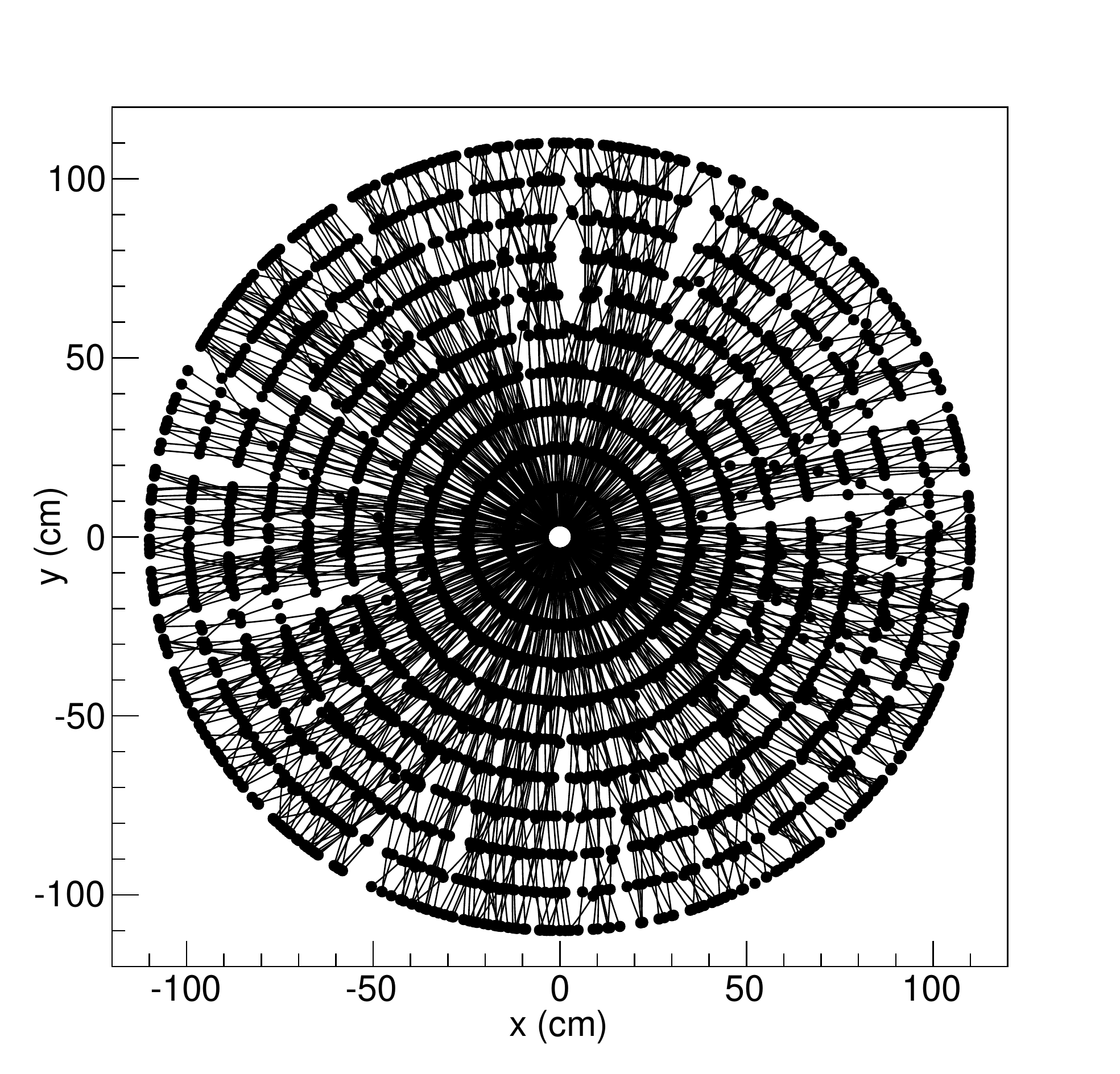} \label{fig:tracks}}
\caption{The Hough transform algorithm applied to a simulation of 500 curved tracks. Left: Hits from Monte Carlo
	simulation for curved tracks originating from the interaction point. Center: The parameter space
	obtained after applying the Hough transform. Right: Final reconstructed tracks. The resulting
	track-finding efficiency for this example is approximately 86\%. From~\cite{Halyo:2013gja}.}
\label{fig:HoughTransform}
\end{center}
\end{figure}

Figure~\ref{fig:HoughTransform} shows an example of the Hough transform algorithm in
action~\cite{Halyo:2013iba}. This simulation uses a simplified detector model, considering only the transverse
plane, with an inner beampipe at radius 3.0~cm, surrounded by ten concentric, evenly-spaced tracking layers
with an outer radius of 110.0~cm. The resolution for a single hit is taken to be 0.4~mm in each direction. In
Figure~\ref{fig:hits}, we see the simulated hits for 500 simulated curved tracks. The Hough transform is then
applied, resulting in the parameter space shown in Figure~\ref{fig:accumulator}. The maxima in the parameter
space are then identified, corresponding to the reconstructed tracks illustrated in
Figure~\ref{fig:tracks}. In this case, the parameter space requires only two parameters (curvature and initial
angle); it should be noted the performance of the Hough transform depends strongly on the number of dimensions
in the parameter space, so a full three-dimensional track reconstruction (requiring five track parameters)
would probably be infeasible. A two-dimensional track reconstruction including displaced tracks would require
a three-dimensional parameter space, which has not yet been implemented and optimized for this study.

\subsection{Performance}

We implemented and benchmarked the Hough transform-based tracking on four different platforms. First, for
comparison purposes, we tested a CPU implementation, using a quad-core Intel i7 3770 CPU with a clock speed of
3.4 GHz, using OpenMP for parallelization. The GPU implementation used an NVIDIA Tesla K20c, which contains
2496 stream processors operating at 706 MHz and 5 GB of on-board memory clocked at 2.6 GHz, and was written in
CUDA C. Finally, the Intel Xeon architecture was tested in two formats: first as a pair of Intel Xeon
E5-2697v2 CPUs, each with 12 physical cores clocked at 2.7 GHz with two-way hyperthreading, and second using
an Intel Xeon Phi QS-7120 coprocessor containing 61 cores at 1.3 GHz, and 16 GB of GDDR5 RAM clocked at 2.8
GHz, using the Intel C++ compiler with the OpenMP framework to produce optimized code.

The code was optimized separately for the CPU, GPU, and Xeon Phi architectures. In particular, the GPU
implementation required special attention to minimize the number of global memory accesses necessary; the
overall event size is relatively small, so that it can be all stored on-board the GPU during processing, so it
is especially important to avoid introducing performance bottlenecks. For the Xeon Phi implementation, the
code was written to enable automatic vectorization as much as possible, avoid cross-thread synchronization
whenever possible and use the best synchronization methods when it was necessary, and to rearrange data access
patterns so that cached data could be reused most efficiently.

\begin{figure}[tbp]
\centering
\includegraphics[width=.7\textwidth]{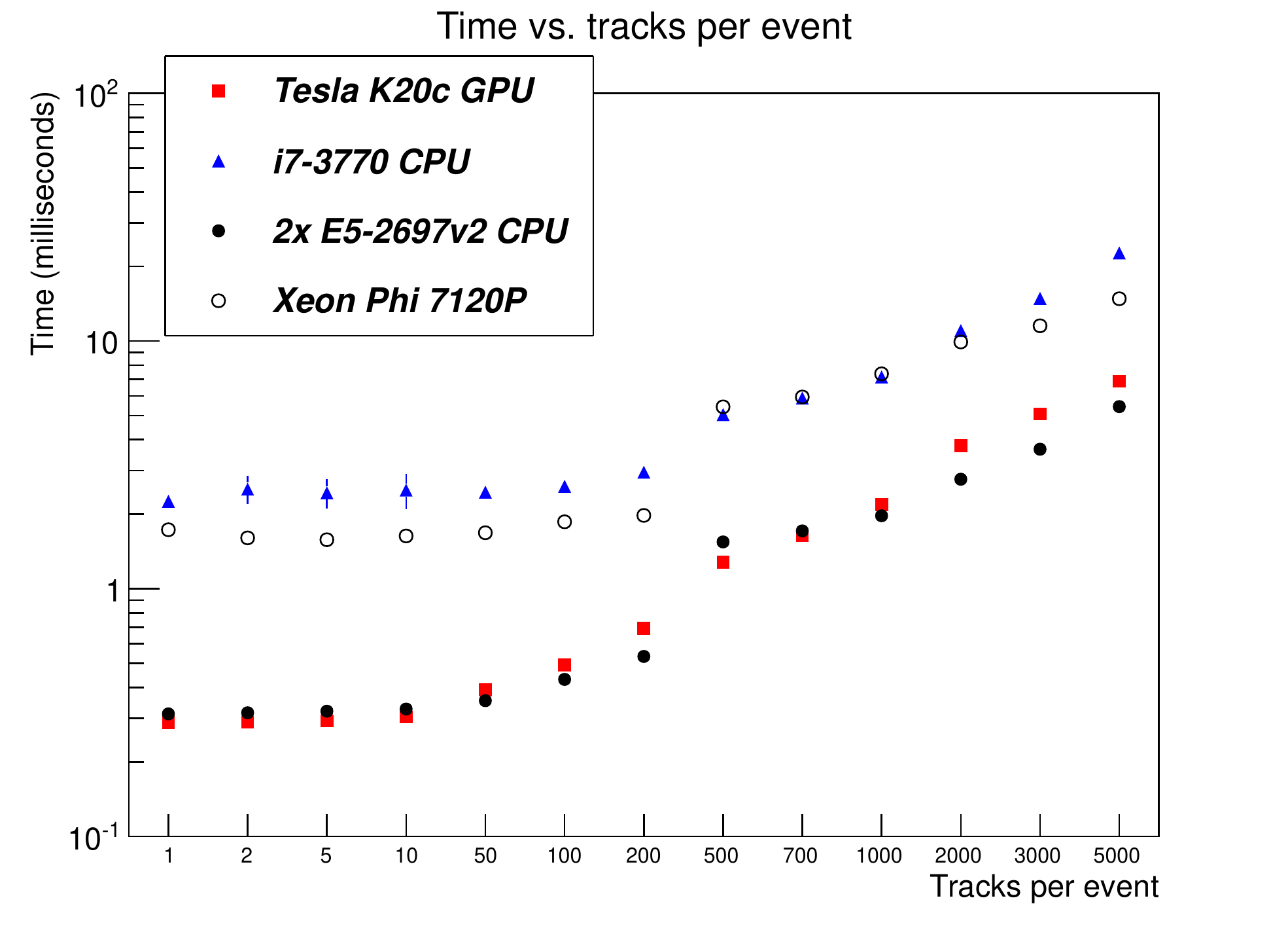}
\caption{Performance of four different implementations of a Hough transform-based tracking algorithm, as a
function of the number of tracks per event. The implementations shown are on an Intel CPU, a NVIDIA Tesla GPU,
an Intel Xeon CPU, and an Intel Xeon Phi coprocessor. From~\cite{Halyo:2013gja}.}
\label{fig:TimePerformance}
\end{figure}

Figure~\ref{fig:TimePerformance} shows the results~\cite{Halyo:2013gja}. We can see that the GPU and Xeon CPU
implementations offer a considerable speedup over the traditional CPU implementation. The Xeon Phi
implementation does not offer a substantial increase; this is due to various features of the problem that are
not as well suited to the MIC architecture. Specifically, when accessing the array representing the parameter
space, the points are not accessed in an orderly fashion, so streaming memory accesses cannot be used; in the
GPU case, this can be mitigated by using the on-chip shared memory, but the Xeon Phi does not have hardware
prefetching for the Level 1 cache and thus is dependent on software prefetching. Future improvements to the
Intel C compiler may alleviate this problem.

In this example, we can see that while parallel computing offers the potential of substantial speedup in the
track reconstruction, the best architecture may depend on the specifics of the problem, and creating an
optimal solution requires attention to a number of details.

\subsection{Displaced Vertex Reconstruction}

\begin{figure}[tbp]
\begin{center}
  \subfigure[]{\includegraphics[width=0.35\textwidth]{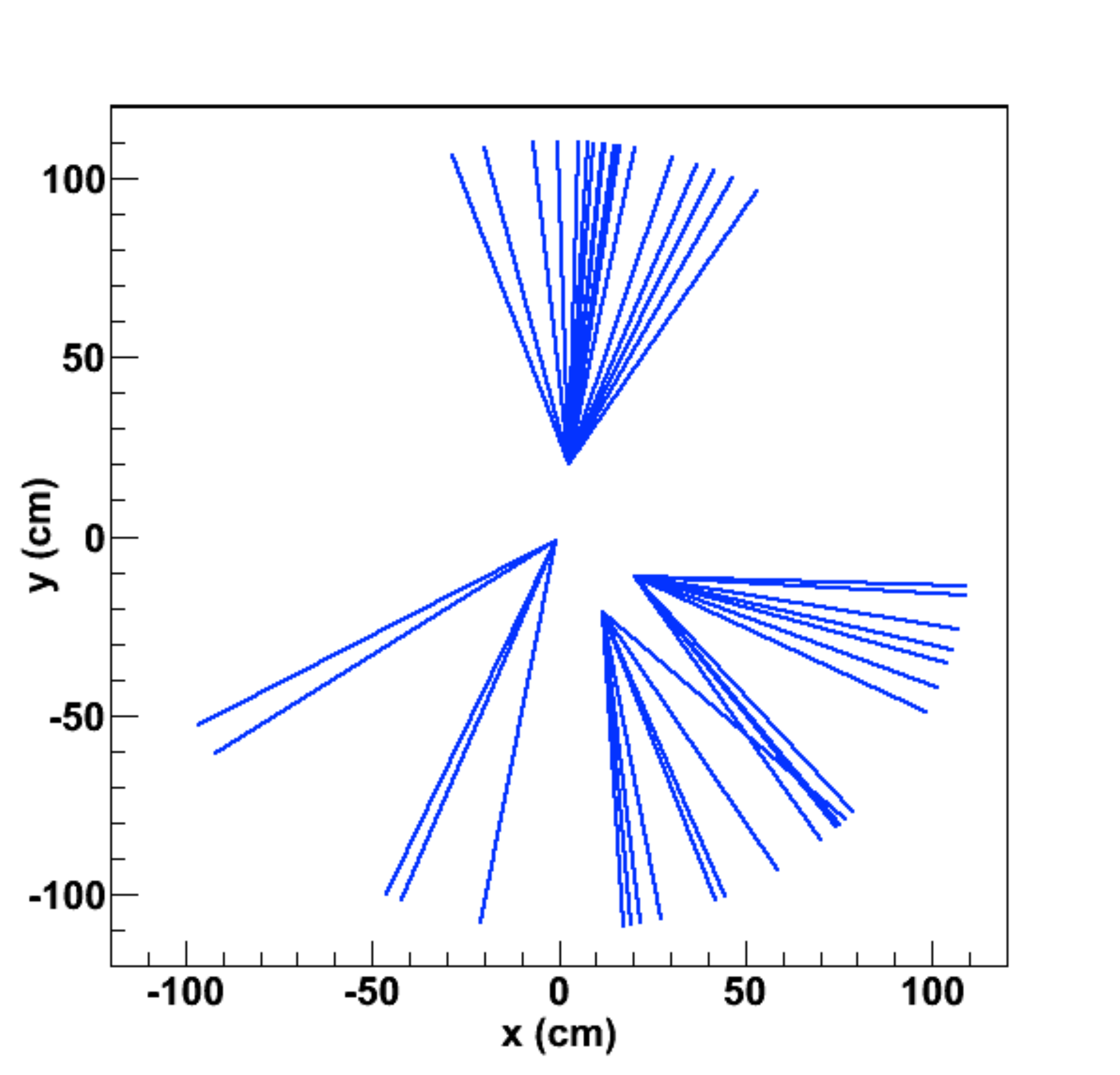} \label{fig:DisplacedJetsTracks}}
  \subfigure[]{\includegraphics[width=0.30\textwidth]{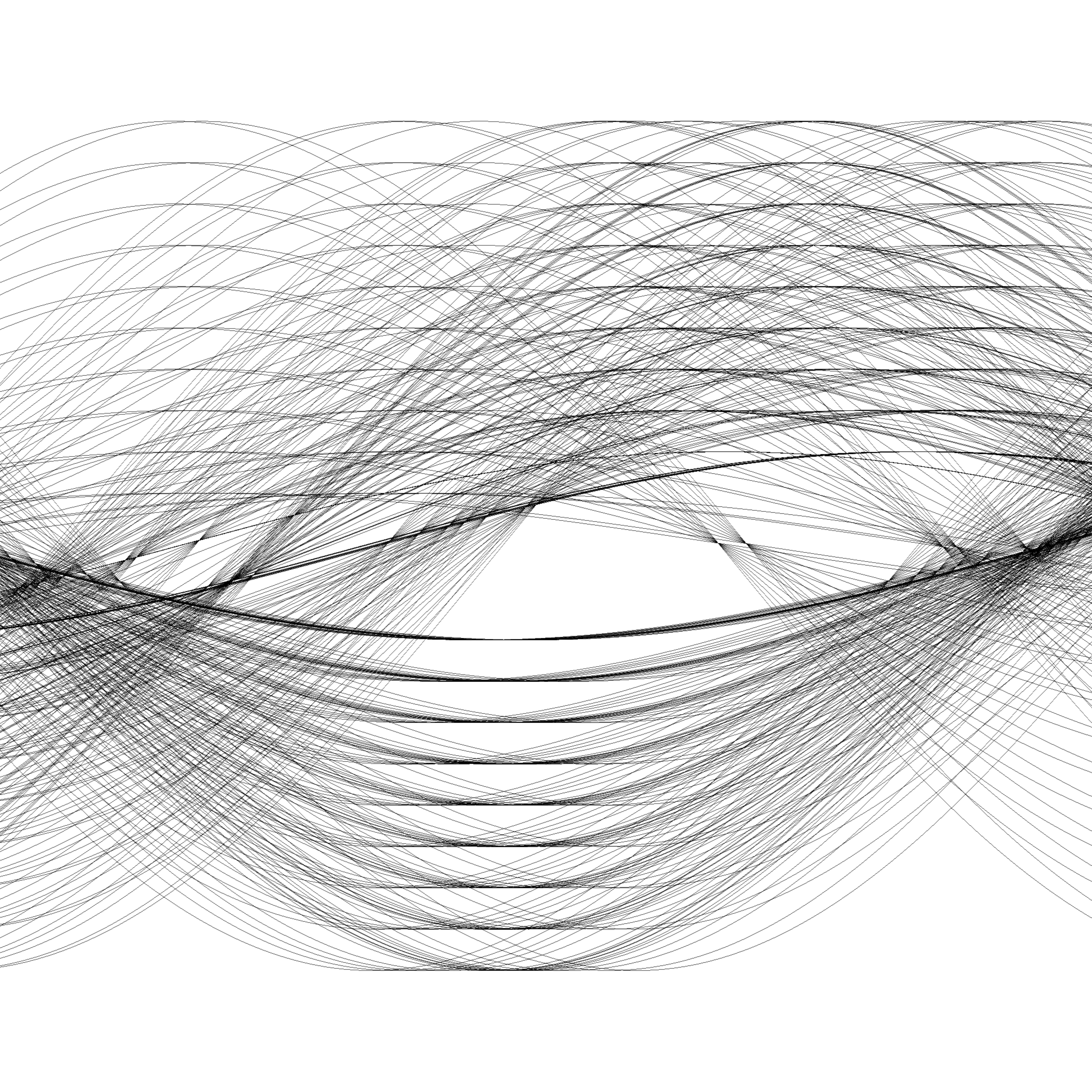} \label{fig:DisplacedJetsAccumulator}}
  \subfigure[]{\includegraphics[width=0.30\textwidth]{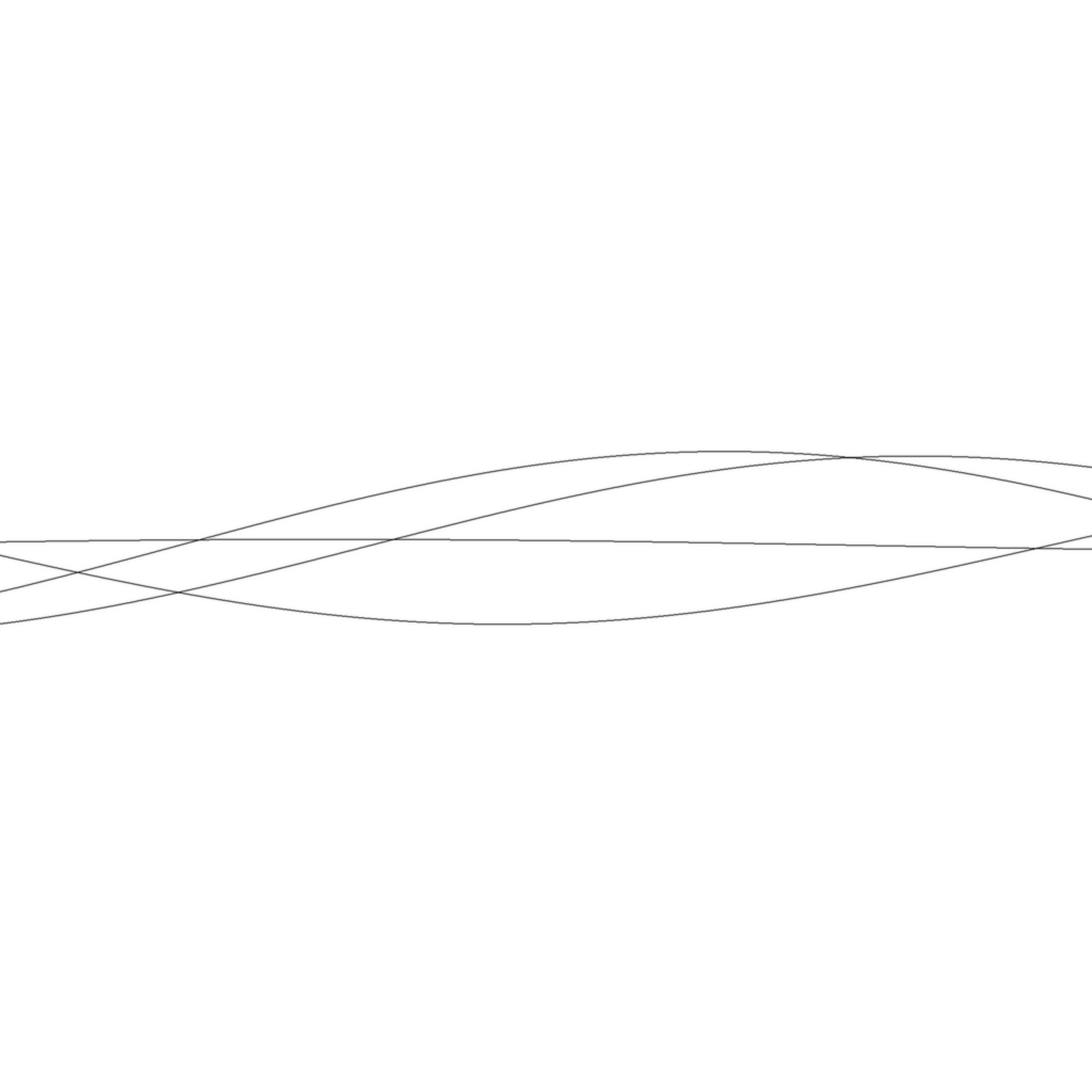} \label{fig:DisplacedJetsVertex}}
  \caption{Hough transform algorithm applied to an event with multiple displaced jets. Left: Simulated tracks
    in the event. Center: The first Hough transform identifies the tracks present in the event. Right: The
    second Hough transform identifies the sinusoids corresponding to the jet vertices. From ~\cite{Halyo:2013cza}.
    \label{fig:DisplacedJets}}
\end{center}
\end{figure}

The Hough transform can do more than reconstruct individual tracks, however. Once the tracks have been
identified, the transform can be run a second time to find displaced vertices by looking for locations where
the tracks intersect. This would be an especially useful capability in searching for events where long-lived
neutral particles decay into jets~\cite{Halyo:2013cza}. Figure~\ref{fig:DisplacedJets} shows an example of
this process. First we simulate tracks for an event with four jets, each produced at a significantly displaced
vertex, shown in Figure~\ref{fig:DisplacedJetsTracks}. Figure~\ref{fig:DisplacedJetsAccumulator} shows the
results of the first Hough transform used to identify the displaced tracks, seen as the maxima in the first
parameter space. Then we can apply a second Hough transform to the result of the first, yielding the result
shown in Figure~\ref{fig:DisplacedJetsVertex}, with the four sinusoids in parameter space corresponding to the
four vertices in the original $r-\phi$ space.

\subsection{Implications for New Physics}

\begin{figure}[tbp]
\centering
\includegraphics[width=.4\textwidth]{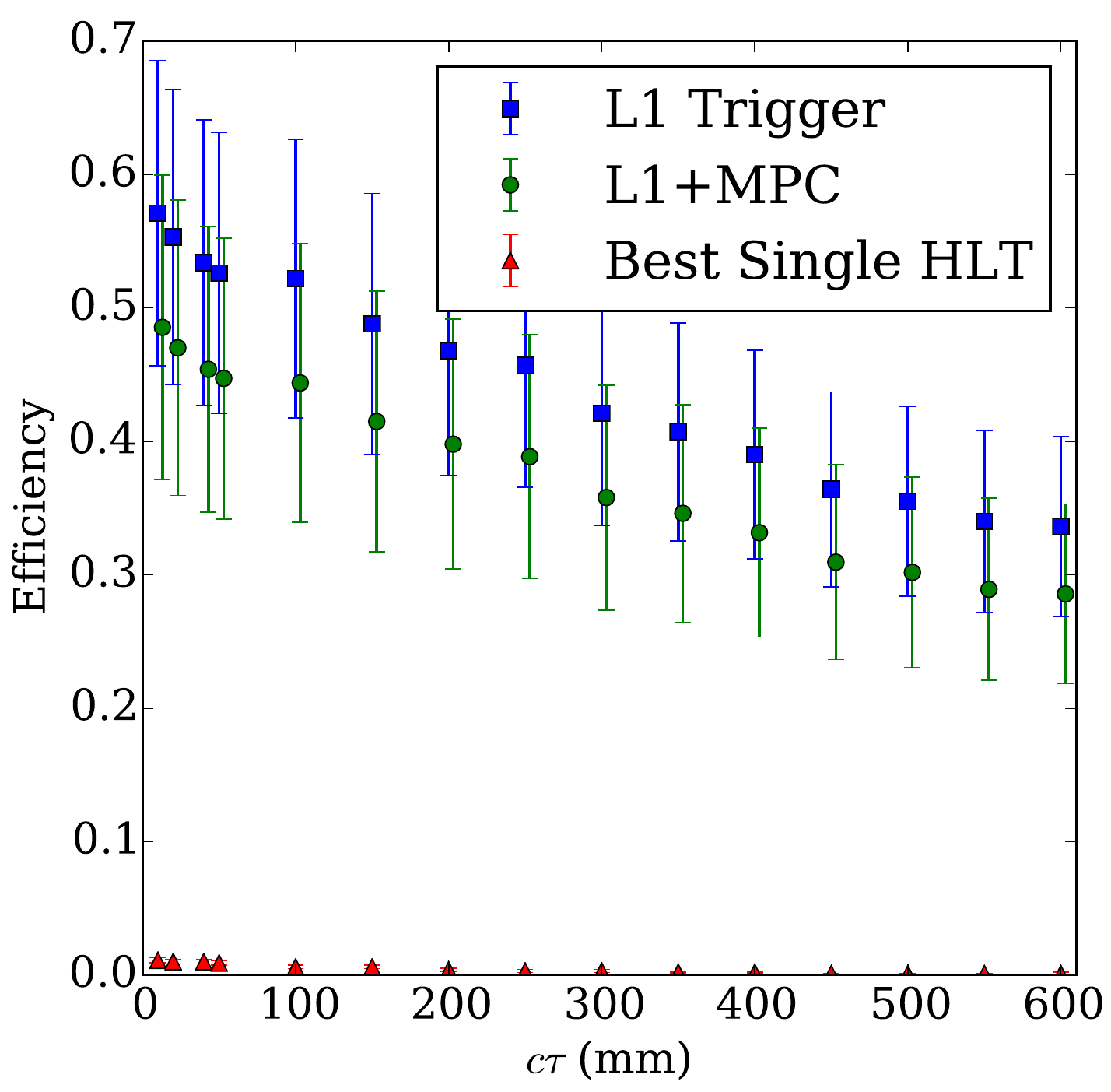}
\includegraphics[width=.4\textwidth]{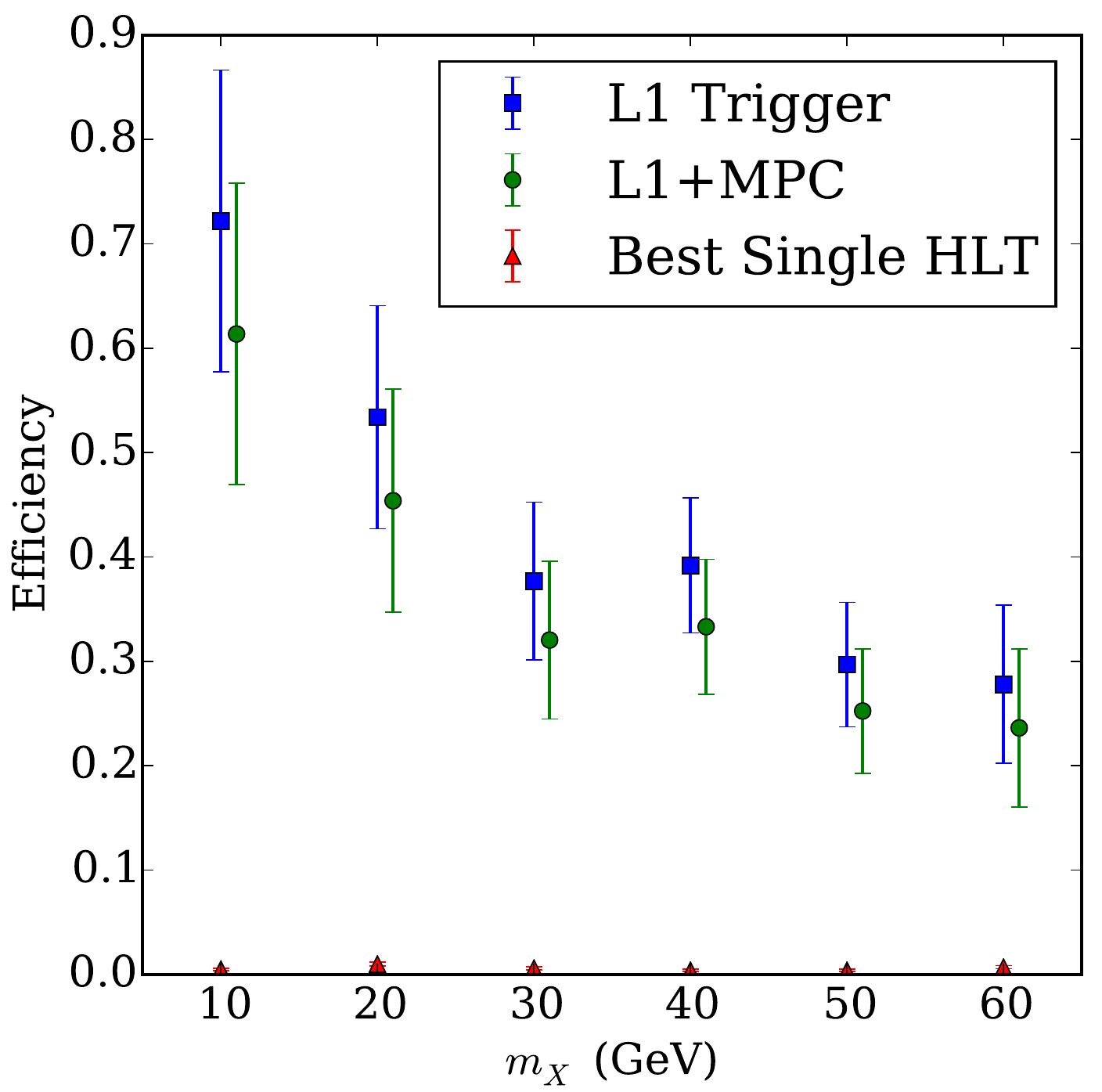}
\caption{Comparison of performance of the CMS L1 trigger (blue), the CMS HLT (red), and a potential MPC-based
trigger (green) for a H $\rightarrow$ XX $\rightarrow b\bar{b}b\bar{b}$ model with $m_H$ = 125 GeV/$c^2$. The
efficiency is shown as a function of the proper lifetime $c\tau$ of the X boson (left) and of the mass of the
X boson $m_X$ (right).}
\label{fig:Higgs4b}
\end{figure}

As an example of how such a trigger could extend the reach to new physics, consider a specific model with a
non-SM Higgs decaying to two long-lived neutral bosons X, each of which decays into a $b\bar{b}$ pair: H
$\rightarrow$ XX $\rightarrow b\bar{b}b\bar{b}$. If $m_{H}$ is small (125 GeV/$c^2$), then such a signature is
very difficult to trigger on; we estimate that the efficiency for the CMS HLT to trigger on such an event
using the 2012 configuration is $<1$\%, and likely to be even smaller given the more stringent trigger
requirements expect for Run 2. However, if we use a MPC-based trigger with an estimated efficiency of 80\%
after the L1 trigger, we estimate that the efficiency would increase to $\sim 30$\%, which would make
discovery possible.~\cite{Buckley:2014ika} Figure~\ref{fig:Higgs4b} shows a comparison between the CMS L1
trigger efficiency, the best trigger efficiency available from the CMS HLT, and the potential efficiency from
a MPC-based trigger.

In this way, we can see that new parallel-based tracking algorithms not only provide the necessary
technological advances to keep up with the increasing quantity and complexity of data as the LHC moves towards
the HL-LHC era, but also opens up the possibility for entirely new avenues of exploration into new physics
areas not accessible currently.

\section{GPU Acceleration in Analysis}

Of course, in addition to the examples discussed above in the online systems, there are also opportunities to
take advantage of GPU acceleration in analyzing the data produced by the LHC. As the quantity of data
increases, and computing-intensive analysis techniques such as matrix element integration methods become more
popular, the benefits of being able to run analyses more efficiently become substantial.

An example of how GPU acceleration can benefit physics analysis is GooFit~\cite{goofit}, a software tool which
provides a parallelized version of the popular RooFit fitting package. GooFit is designed so that the end user
can use it in much the same manner as RooFit; in most cases, no specific knowledge of GPU programming is
necessary, unless the user wishes to implement their own custom probability density functions. GooFit supports
two separate backends for running on different parallel architectures, one written in CUDA and one using
OpenMP.

To demonstrate the power of GooFit, the authors used it in a time-dependent Dalitz-plot analysis for the decay
$D^0 \rightarrow \pi^+\pi^-\pi^0$, with a total of approximately 40 free parameters in the unbinned fit in a
dataset of approximately one hundred thousand events. Using an NVIDIA Tesla C2050, the multithreaded GPU
implementation achieves a speedup factor of ~300 compared to the original unoptimized, single-threaded CPU
implementation and is approximately 5 times faster than a fully-optimized multithreaded CPU version (running
16 threads on 8 cores). Even on a laptop with an ordinary consumer graphics card (NVIDIA GT 640M), the GPU
implementation achieves a speedup factor of ~90 over the original.~\cite{goofit}

\section{Conclusions}

As the LHC continues to reach new heights of performance, it is necessary for the computing supporting the LHC
efforts to similarly advance in order to process the increasing amounts of data. Parallel computing techniques
provide a way to enhance performance, often dramatically, so that the computing will not only be able to keep
pace, but also provide the capabilities necessary to extend the physics reach of the detectors and continue
the search for new physics at the LHC. While some gains can be realized with relatively simple software
solutions, obtaining the best performance will often require careful analysis and optimization in order to
best use the features of GPU and/or MIC architecture. On the other hand, GPU and MIC accelerators offer the
benefit of being easy to add to existing computing systems. Clearly, much work remains in order to fully
integrate MPC technologies into the LHC experiments, but early results have shown a great deal of promise, and
strongly suggest that MPC technologies will need to be an important part of the future of the LHC.

\acknowledgments

The authors would like to thank the organizers of the INFIERI Summer School, especially Aurore Savoy-Navarro,
for their work in making this excellent school happen. This work is supported by the US Department of Energy,
Office of Science Early Career Research Program under Award Number DE-SC0003925.

\end{document}